\documentclass[aps,twocolumn,preprintnumbers,showpacs,showkeys]{revtex4}
\usepackage{epsfig}
\usepackage{amssymb,amsmath,amsfonts,amsthm,graphicx}
\usepackage{hyperref}
\usepackage{cleveref}
\graphicspath{{./Figures/}}                 
\newcommand{\noi}{\noindent}
\newcommand{\beq}{\begin{equation}}
\newcommand{\eeq}{\end{equation}}
\newcommand{\bea}{\begin{array}}
\newcommand{\eea}{\end{array}}
\newcommand{\beqa}{\begin{eqnarray}}
\newcommand{\eeqa}{\end{eqnarray}}
\newcommand{\Fig}[1]{Fig.~\ref{#1}}

\def\beqa{\begin{eqnarray}}
\def\eeqa{\end{eqnarray}}

\begin{document}

\preprint{ITEP-LAT/2017}

\title{Dyons near the transition temperature in $SU(3)$ lattice gluodynamics \\}

\author{V.~G.~Bornyakov}
\affiliation{Institute for High Energy Physics NRC ``Kurchatov Institute'',
142281 Protvino, Russia, \\
Institute of Theoretical and Experimental Physics, 117259 Moscow, 
Russia \\
School of Biomedicine, Far East Federal University, 690950 Vladivostok, 
Russia}

\author{E.-M. Ilgenfritz}
\affiliation{Joint Institute for Nuclear Research, BLTP, 141980 Dubna, Russia}

\author{B.~V.~Martemyanov}
\affiliation{Institute of Theoretical and Experimental Physics, 117259 Moscow, 
Russia \\
National Research Nuclear University MEPhI, 115409, Moscow, Russia \\
Moscow Institute of Physics and Technology, 141700, Dolgoprudny, Moscow Region,
Russia}
\date{\today}
\begin{abstract}
We study the topological structure of $SU(3)$ lattice gluodynamics by cluster
analysis. This methodological study is meant as preparation for full QCD.
The topological charge density is becoming visible in the process of
overimproved gradient flow, which is monitored by means of the the Inverse 
Participation Ratio (IPR). The flow is stopped at the moment when calorons 
dissociate into dyons due to the overimproved character of the underlying 
action. This gives the possibility to simultaneously detect all three dyonic 
constituents of KvBLL calorons in the gluonic field.
The behaviour of the average Polyakov loop under (overimproved) gradient
flow could be also (as its value) a diagnostics for the actual phase the configuration is belonging
to.
Timelike Abelian monopole currents and specific patterns of the local Polyakov
loop are correlated with the topological clusters.

The spectrum of reconstructed cluster charges $Q_{cl}$ corresponds
to the phases. It is scattered around $Q_{cl} \approx \pm 1/3$ in the
confined phase, whereas it is $Q_{cl} \approx \pm 0.5 \div 0.7$ for heavy
dyons and $|Q_{cl}| < 0.3$ for light dyons in the
deconfined phase. 
We estimate the density of heavy and light dyons at three values of temperature. 
We find that heavy dyons are increasingly
suppressed with increasing temperature.

The paper is dedicated to the memory of Michael M\"uller-Preussker who was
a member of our research group for more than twenty years.

\end{abstract}

\keywords{Lattice gauge theory,  caloron, dyon}

\pacs{11.15.Ha, 12.38.Gc, 12.38.Aw}

\maketitle

\section{Introduction}
\label{sec:introduction}
There are topological objects in the vacuum of Yang Mills theory and QCD 
(at finite temperature) known as constituent ``dyons'' of 
Kraan-van Baal-Lee-Lu 
(KvBLL) calorons~\cite{Kraan:1998pm,Kraan:1998sn,Lee:1998bb}. 
These objects have been studied  in $SU(2)$ and $SU(3)$ 
lattice gluodynamics and then in lattice QCD both in the confined and deconfined phases
~\cite{Bornyakov:2007fm,Bornyakov:2008im,Ilgenfritz:2013oda,Bornyakov:2014esa,Bornyakov:2015xao}. 
At the beginning one or the other method of smoothing of the lattice gauge field to get 
rid of the ultraviolet fluctuations was used in these studies.
The fermionic topological charge density constructed from low lying modes 
of the overlap Dirac operator (for two or three types of temporal boundary 
conditions imposed on the fermion field) was more recently used for this goal.  
Methods connected to the spectrum of the overlap Dirac operator are rather 
computer time consuming. For this reason  we recollect here, first for the example 
of pure gluodynamics, how much information in this respect can be obtained 
by means of purely gluonic observables: the gluonic topological charge density,
the Polyakov loop (global and local) and Abelian monopoles (obtained in the 
Maximal Abelian Gauge \cite{Kronfeld:1987vd}). Many of these tools have been proposed and used by us
occasionally in the past.

The main problem with the gluonic topological charge density is that it is 
too noisy when measured on the vacuum lattice gluonic fields 
(in Monte Carlo or Hybrid Monte Carlo ensembles) such 
that a topological structure is not discernible. 
The objects we are looking for (calorons, dissociated or not dissociated 
into dyons) appear only in the process of cooling or its continuous analog 
- the gradient flow~\cite{Luscher:2009eq,Luscher:2010iy,Luscher:2011bx}, 
usually realized with respect to the Wilson action (therefore Wilson flow). 
The gradient flow with respect to so-called over-improved 
action~\cite{GarciaPerez:1993ki} has an additional built-in feature of 
forced dissociation of calorons into dyons~\cite{Bruckmann:2004ib}. 
This is an advantage if we want to analyze the gauge field configurations 
down to the caloron-constituents - the dyons. 
During this process, the dissociation of calorons can be controlled by 
monitoring the inverse participation ratio (${\mathrm{IPR}}$) of the modulus 
$|q(x)|$ of the topological charge density. This IPR varies between the 
extreme values: ${\mathrm{IPR}}=1$ (for a homogeneously delocalized density) 
and ${\mathrm{IPR}}=V_4$ (for a pointlike localized density). $V_4$ is the four 
dimensional volume. The IPR is defined (analogously to the IPR of the scalar 
density $|\psi(x)|^2$ of fermionic modes) as follows:
\beq
{\mathrm{IPR}} = V_4 \frac{\sum_x |q(x)|^2}{(\sum_x |q(x)|)^2}~~.
\label{eq:localization}
\eeq
The IPR increases  when ultraviolet fluctuations of the gluonic field are 
removed. It  also increases when already existing topological structures 
are removed by annihilation 
e.g. calorons and anticalorons as well as their dyon and antidyon 
constituents. At the same time the dissociation of calorons into dyons leads to a 
decrease of the IPR. So, we are in the position to stop the process of 
over-improved gradient flow (monitoring the IPR) when calorons are maximally 
delocalized (dissociated to dyons). We found that, as a rule, the IPR of a given configuration
first grows to a maximum. Then it decreases before it grows again.

  We expect that maximal delocalization (dissociation) of the calorons
should happen (approximately) at the first minimum 
after the first maximum of the ${\mathrm{IPR}}$ in the over-improved gradient 
flow history. 
In practice we used a fixed flow time for all flow histories 
(at the same temperature) when they typically go through the above mentioned 
minimum.

Let us note that the dyons, or quark instantons, are playing the decisive
role in a recent model of the QCD vacuum proposed by
Shuryak and collaborators \cite{Shuryak:2011aa,Faccioli:2013ja,Larsen:2015vaa,Larsen:2014yya, Larsen:2017sqm}. This model serves to expain the intermediate temperature dependence of the topological suceptibility before
the onset of the Dilute Gas Approximation (DIGA).
The lattice computation of the density of dyons is important for a cross-check
of this model.

\section{Thermal Ensembles}
\label{sec:thermal-ensembles}

For this methodical study of dyon detection in $SU(3)$ gauge theory we employ 
quenched ensembles which are generated with the standard Wilson 
action $S_W$  for the lattice coupling $\beta = 6/g_0^2 \,$.

To fix the corresponding lattice spacing $a$ as a function of $\beta$ for 
this action we rely on the Necco--Sommer parametrization \cite{Necco:2001gh}.

The test ensembles (to describe three temperatures)  were generated on the asymmetric 
lattices with a four-dimensional volume $V_4 = a^4 L_t\cdot L_s^3$, where 
$L_t=6$ is the number of sites in the temporal direction, and $L_s=24$ 
the size in all spatial directions.
The phase transition for $L_t=6$ takes place at 
$\beta_c = 5.894$ \cite{Iwasaki:1992ik}. It corresponds to a critical 
temperature of pure gluodynamics $T_c \simeq 300$ MeV \cite{Gattringer:2002mr}.

Our test ensembles (100 configurations each ) for the three temperatures of 
interest, $T=0.79~T_c$, $T=1.27~T_c$ and $T=1.5~T_c$, were generated with 
$\beta=5.8$, $\beta=6.0$ and $\beta=6.084$, respectively.

\section{Flow histories}
\begin{figure*}[!htb]
\centering
\hspace{0.3cm}\includegraphics[width=.7\textwidth]{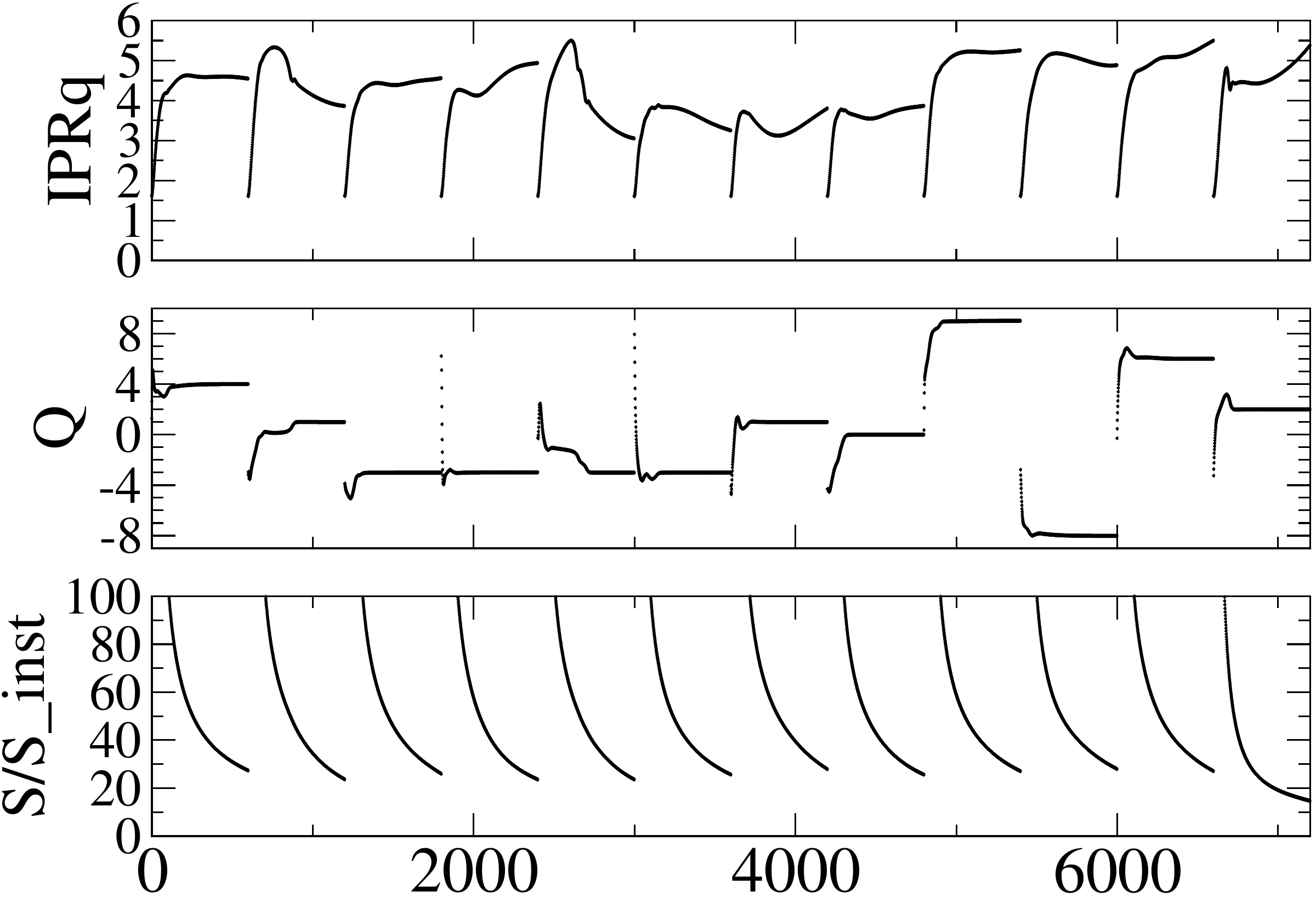}%
\hspace{0.3cm}
\caption{Over-improved flow histories for 12 configurations below $T_c$ 
($T= 0.79~T_c$ ) shown 
 for ${\mathrm{IPR}}$, the topological charge $Q$ and the action in 
instanton units (from top to bottom). On X-axes the cumulative number of flow steps is shown 
(for every configuration 600 flow steps). }
\label{histories1c}
\end{figure*}

\begin{figure*}[!htb]
\centering
\hspace{0.3cm}\includegraphics[width=.7\textwidth]{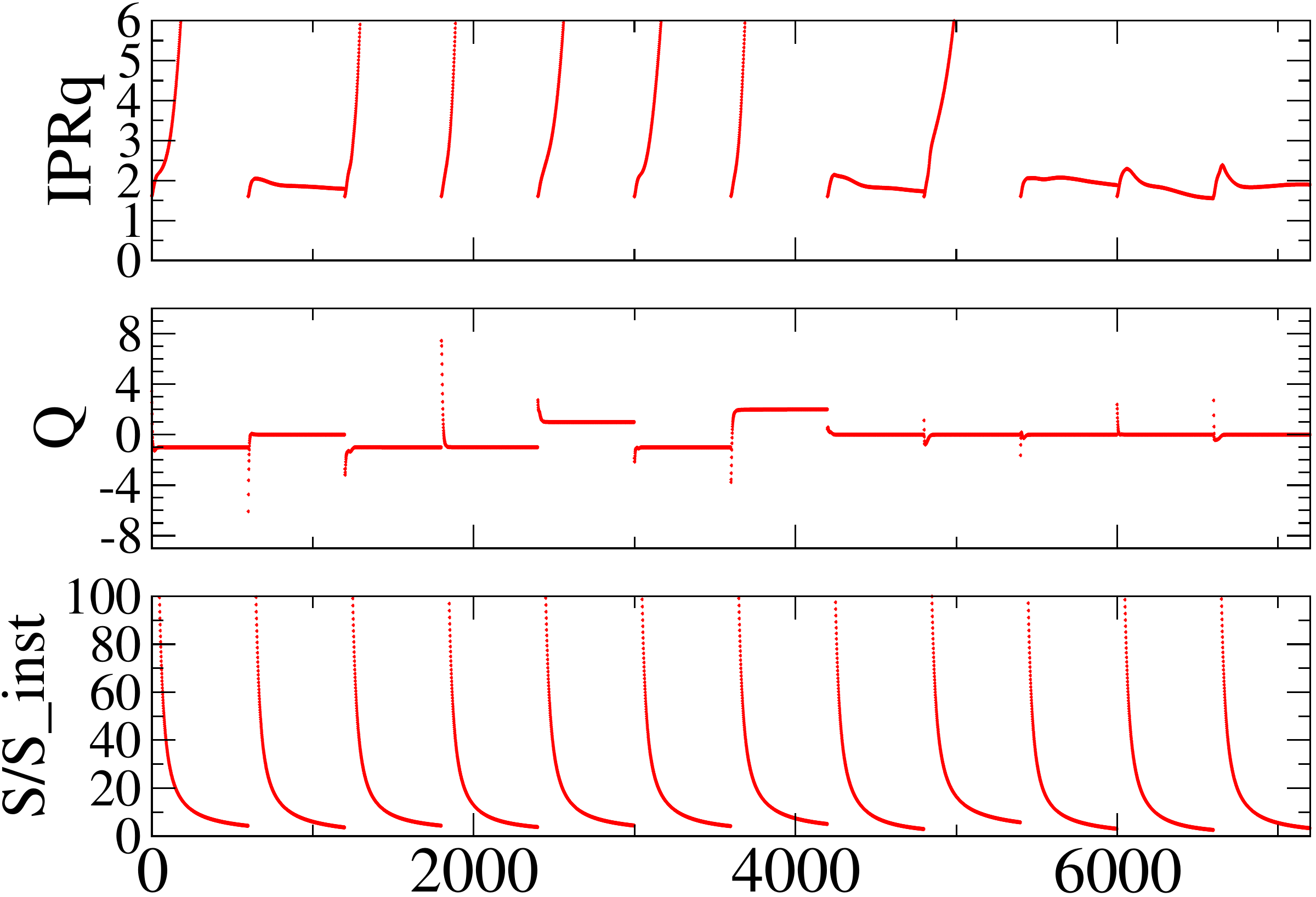}%
\hspace{0.3cm}
\caption{Over-improved flow histories for 12 configurations above $T_c$ 
($T= 1.27~T_c$ ) shown 
 for ${\mathrm{IPR}}$, the topological charge $Q$ and the action in 
instanton units (from top to bottom). On X-axes the cumulative number of flow steps is shown 
(for every configuration 600 flow steps).}
\label{histories1d}
\end{figure*}

\begin{figure*}[!htb]
\centering
a)\hspace{0.3cm}\includegraphics[width=.45\textwidth]{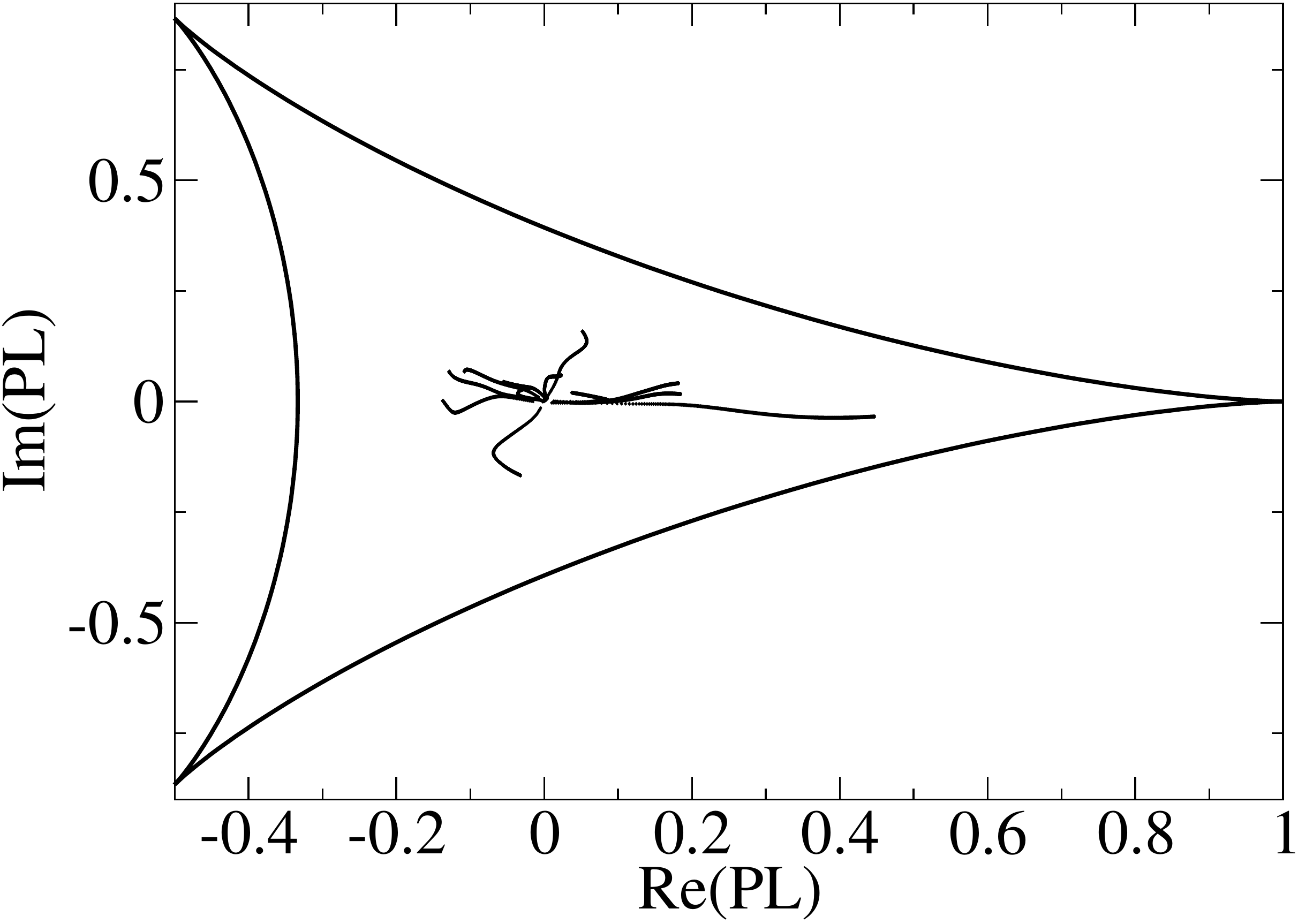}%
\hspace{0.3cm}
b)\hspace{0.3cm}\includegraphics[width=.45\textwidth]{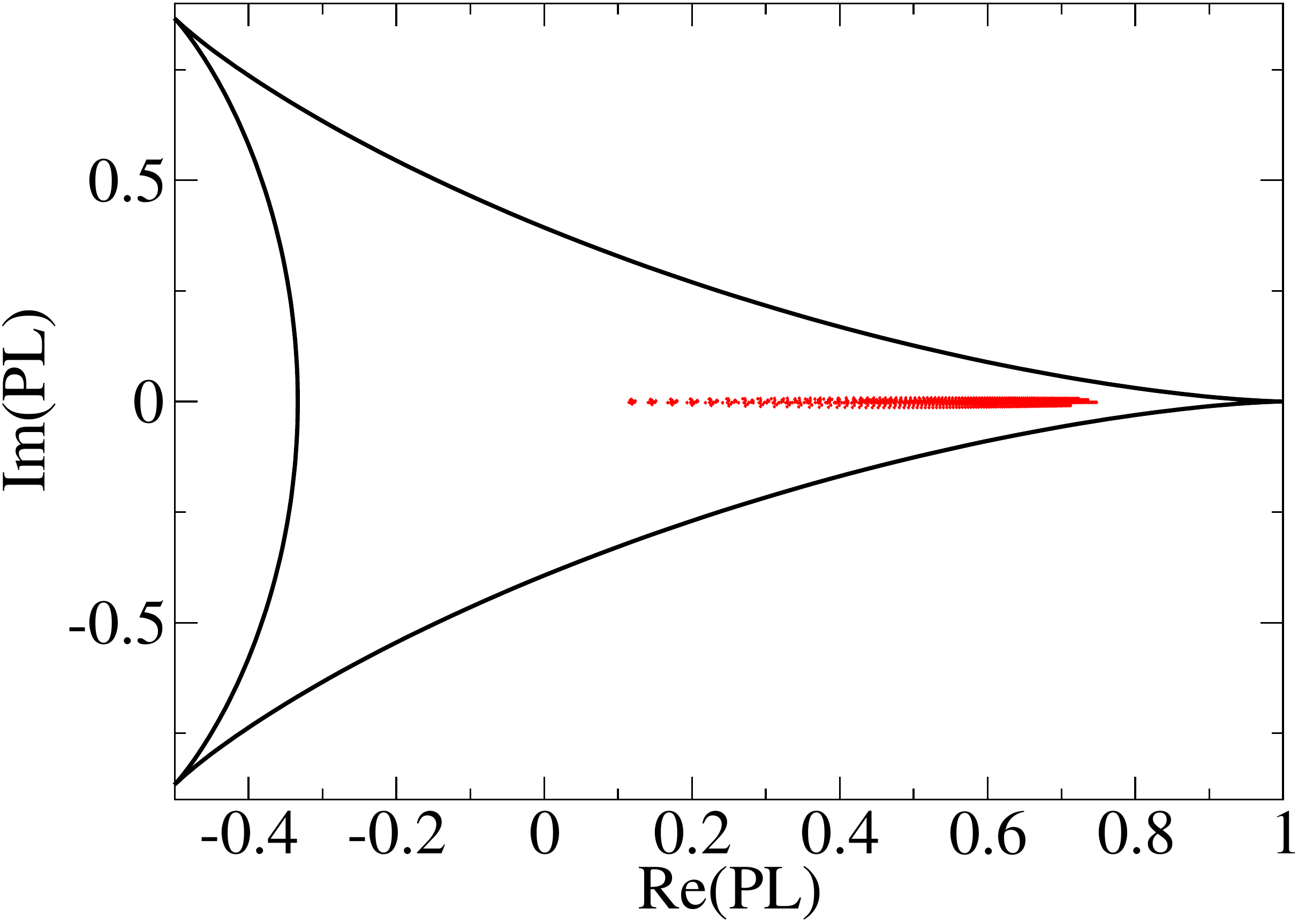}\\
\vspace{1cm}
\caption{Over-improved flow histories of the volume-averaged Polyakov loop
for 12 configurations below/above $T_c$: 
(a) at $T= 0.79~T_c$ and (b) at $T= 1.27~T_c$, respectively.}
\label{histories2}
\end{figure*}

 For illustration, over-improved gradient flow histories 
for 12 configurations 
taken out of two ensembles,
below and above $T_c$ ($T= 0.79~T_c$
and $T= 1.27~T_c$),
 are shown
in \Fig{histories1c} and \Fig{histories1d},
respectively. 
The figures show 
in parallel the localization of the topological density ${\mathrm{IPR}}$, 
the total topological charge $Q$, and the action in instanton action units (from 
top to bottom) Additionally in \Fig{histories2} the flow histories of the volume-averaged Polyakov loop
(PL) are presented. In all cases 600 steps (with step length $\epsilon = 0.02$) of flow 
are shown. 


The flow histories for the volume-averaged Polyakov loop are different 
in the confining and deconfining phases: while in the confining phase PL has 
no prefered direction of evolution (interpreted in \cite{Bornyakov:2013iva} 
as indication of an equal number of dyons of all three types), in the 
deconfining phase PL goes in the direction where rare heavy dyons -- left from 
supposedly asymmetric caloron solutions -- become more and more heavy, whereas 
more abundant light dyons become more and more light (also this interpretation 
was given in \cite{Bornyakov:2013iva}). 

The flow histories for the topological charge show very fast stabilization 
towards integer topological charges ( up to $Q = \pm 8$) 
in the confining phase and towards $Q = 0$ or $Q = \pm 1$ in the deconfining 
phase. In the confining phase the characteristic flow history for ${\mathrm{IPR}}$  
is as follows: ${\mathrm{IPR}}$s first grows, then decreases
until -- at several hundreds of flow steps -- it goes through a local minimum,
before finally it grows again. 
In the deconfining phase, ${\mathrm{IPR}}$s always monotonously grows, 
 for $Q = \pm 1$. For $Q = 0$ configurations, the ${\mathrm{IPR}}$ 
typically experiences no sizable changes under gradient flow. 
In all cases, the action monotonously decreases, although it most rapidly 
does so at $T > T_c$. 

Like in our previous paper \cite{Bornyakov:2013iva}, we will interpret also 
the ${\mathrm{IPR}}$ flow histories in terms of a dyonic picture. The growth 
of ${\mathrm{IPR}}$ in the confined phase can be understood as result of two processes: the 
removal of perturbative fluctuations and of the decreasing of the number of separate 
topological objects in the process of annihilation of calorons and anticalorons
(as well as of their constituent dyons and antidyons). 
The topological charge density becomes more localized when perturbative 
fluctuations are hidden under the peaks of the topological charge density 
of topological objects.
The subsequent drop of ${\mathrm{IPR}}$ we may interpret as the result of an increasing of 
 the number of separate topological objects in the process of 
dissociation of calorons into constituent dyons. Thus, the first minimum 
(after passing the first maximum) in the history curve of ${\mathrm{IPR}}$  
can be considered as a point of maximal dissociation of calorons into dyons 
in the course of over-improved gradient flow. We believe, it is a good choice
to stop the flow process here for the observation of these dyons. As it can 
be seen from \Fig{histories1c} for ${\mathrm{IPR}}$, in the confined phase 
several hundreds of flow steps is needed to reach this point.  
In practice we used 600 steps to be sure that maximal dissociation has
been achieved for all confining configurations.

In the deconfined phase, the topological objects expected are heavy dyons 
with highly localized and high-valued topological charge density on one hand
and light dyons with delocalized and low-valued topological charge density
on the other. The calorons are expected to be already dissociated into these 
asymmetric kinds of dyons. For $Q = 0$ configurations in the deconfined phase
we have no heavy dyons, and the IPR will experience
no sizable changes during the flow process.
For $Q = \pm 1$, the final ${\mathrm{IPR}}$ is defined (after removal of
perturbative fluctuations) by a single heavy dyon with highly localized 
charge density and its absolute topological charge being close to one. 

Since we have no stopping criterium for the deconfined phase that would be
motivated by the ${\mathrm{IPR}}$ history (opposite to the confined phase) 
we finish the flow process after approximately 100 steps of gradient flow  
when the action of a configurations is comparable with that after 600 steps 
of flow in the confined phase.

\section{Correlation between Abelian monopoles and Polyakov loops in 
after-flow gluon fields}

Relatively isolated dyons in KvBLL caloron solutions are monopoles, and 
the local holonomies (untraced Polyakov loops) have the specific property 
that two eigenvalues of the holonomy become degenerate there \cite{VanBaal:2001pm}. 
In the case of $SU(2)$ gauge theory this corresponds to the points where 
Polyakov loops take the values $\pm I$. 
Abelian monopoles (in the sense of Abelian projection after fixing MAG) 
and dyon constituents of KvBLL caloron solution (placed on the lattice) 
are in one-to-one correspondence (at least for isolated objects), and 
in the $SU(2)$ case they are correlated with the above mentioned 
points~\cite{Ilgenfritz:2004zz,Ilgenfritz:2006ju}.

In the $SU(3)$ case the degeneracy of Polyakov loops corresponds to the 
flanks of the unitarity triangle~\cite{VanBaal:2001pm}. For a Polyakov loop 
$$\mathrm{diag}(\mathrm{e}^{2\pi i\mu_1},\mathrm{e}^{2\pi i\mu_2},
\mathrm{e}^{2\pi i\mu_3})$$
(with $\mu_1 \leq \mu_2 \leq \mu_3 \leq \mu_4 = 1+\mu_1$
and $\mu_1+\mu_2+\mu_3 = 0$) parametrised by three numbers
$m_1 = \mu_2 - \mu_1$, $m_2 = \mu_3 - \mu_2$ and $m_3 = \mu_4 - \mu_3$
this degeneracy happens when either $m_1$ or $m_2$ or $m_3$ vanish.

\begin{figure*}[!htb]
\centering
a)\hspace{0.3cm}\includegraphics[width=.45\textwidth]{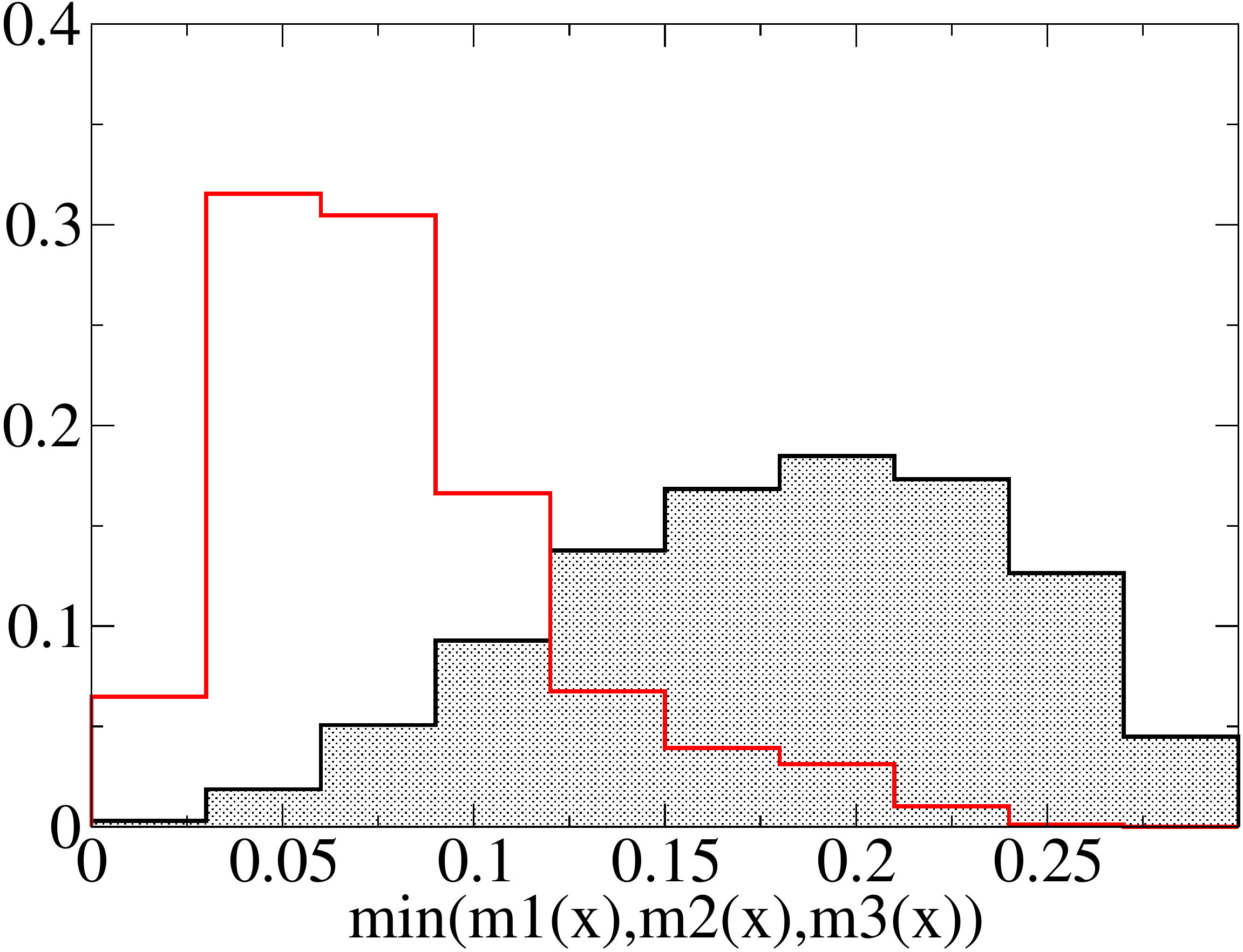}%
\hspace{0.3cm}
b)\hspace{0.3cm}\includegraphics[width=.45\textwidth]{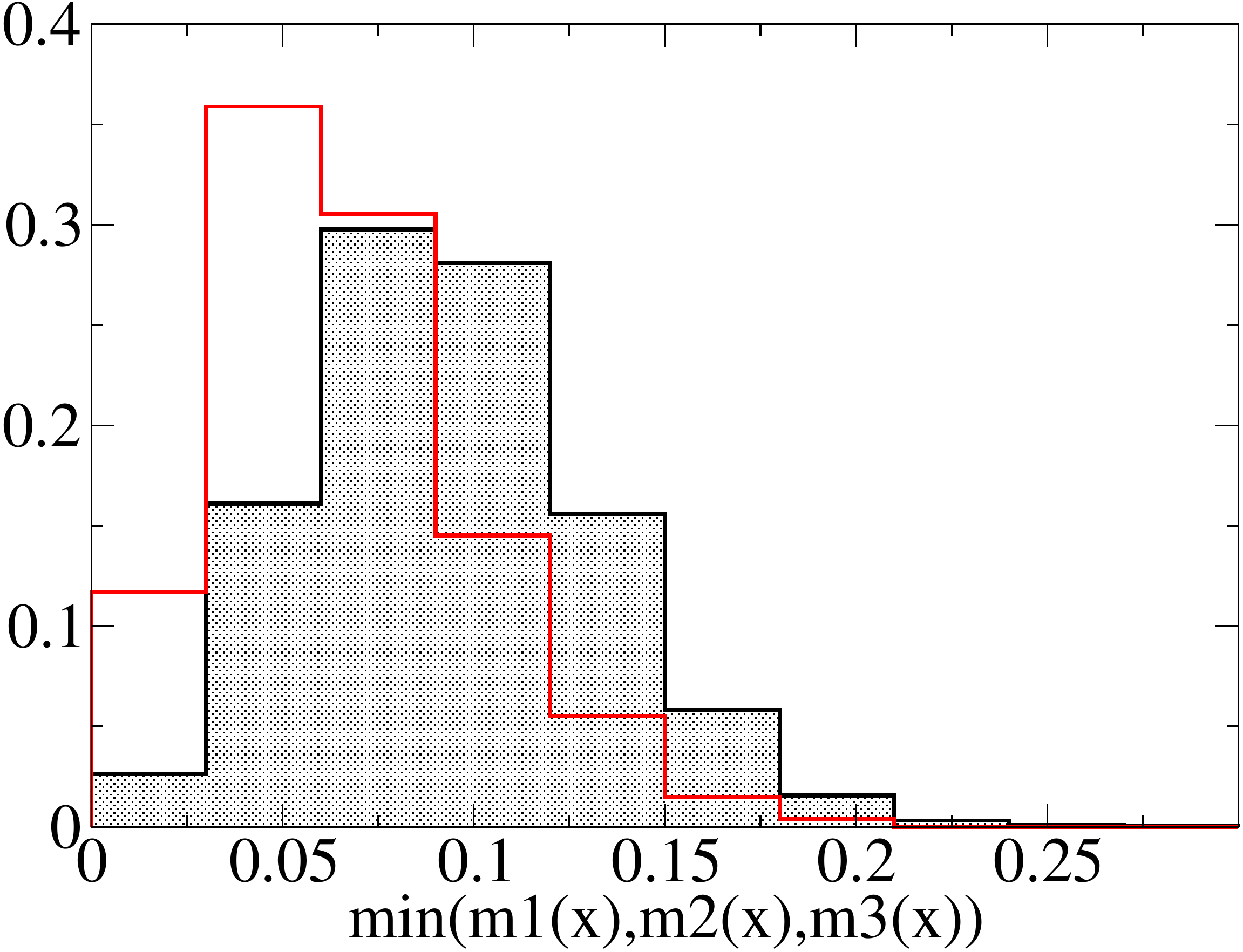}\\
\caption{The distributions with respect to the minimal distance 
$\rm{min}(m_1(x), m_2(x), m_3(x))$ between the local Polyakov loop value 
and the nearest flank of the Polyakov triangle over all lattice sites 
(shaded histogram) and for all cubes where thermal monopoles are located 
(open red histogram) are shown in (a) for $T= 0.79~T_c$ in the confined 
phase and in (b) for $T= 1.27~T_c$ in the deconfined phase.} 
\label{pl_versus_monopoles}
\end{figure*}

The inter-correlation of Abelian monopoles and the monopoles defined by 
degenerated eigenvalues of holonomy  can be detected by  
a correlation between loci of minimum of $m_1$, $m_2$, $m_3$ with 
Abelian monopoles localized in MAG~\cite{Bornyakov:2015xao}.
Which one of $m_1$, $m_2$ or $m_3$ becomes very small determines the
type of dyon. In \Fig{pl_versus_monopoles} the distributions of 
$\rm{min}(m_1(x), m_2(x), m_3(x))$ taken over all lattice sites (shaded 
histogram) is compared with the distribution over all cubes which are 
duals of time-like monopole links (open red histogram).  
The latter are loci where thermal monopoles are located.
This is shown for $T= 0.79~T_c$) in the confined phase (Fig. 4 a)
and for $T= 1.27~T_c$ in the deconfined phase (Fig. 4 b).
In the case of a monopole the minimum $\rm{min}(m_1(x), m_2(x), m_3(x))$ 
is taken over the eight corners of the three-dimensional cube containing 
that monopole. It is seen that the distibution of 
$\rm{min}(m_1(x), m_2(x), m_3(x))$ on thermal monopoles is shifted towards 
zero compared with the distribution in the bulk.
It is also seen that in the deconfined phase, when the Polyakov loop
averaged over volume is near the corner
of the unitarity triangle ((1,0) in our case) 
the distibution of $\rm{min}(m_1(x), m_2(x), m_3(x))$ on thermal monopoles 
which now correspond to light dyons has only minor  shift in compariosn  
with the distribution in the bulk.

\begin{table*}[ht]
\begin{center}
\vspace*{0.5cm}
\begin{tabular}{|l|l|l|c|c|c|c|c|c|}
\hline
 phase&$V_{cl}$ & $V_{clmon}$ & $ N_{cl} $ & $N_{clmon}$ & $ N_{mon} $ & $N_{moncl}$
  & $ N_{loop} $ & $N_{loopcl}$ \\
\hline
$T= 0.79 ~T_c < T_c$&$2.6(1)\%$ & $2.4(1)\%$ & $ 15.4(2) $ & $10.4(2)$ &
$ 306(6) $ & $94(3)$& $ 45(1) $ & $ 19(1)$ \\
\hline
$T= 1.27 ~T_c> T_c$&$2.7(3)\%$ & $1.0(1)\%$ & $29(2) $ & $3.0(2)$ &
$ 130(3) $ & $21(2)$& $ 21(1) $ & $ 5.2(4)$ \\
\hline
$T= 1.5 ~T_c> T_c$&$4.7(3)\%$ & $1.2(1)\%$ & $ 45(2) $ & $3.0(2)$ &
$ 106(3) $ & $19(1)$& $ 18(1) $ & $ 5.9(4)$ \\
 \hline
\end{tabular}
 
\label{tabdata}
\vspace*{0.5cm}
\caption{Results of the cluster analysis.
All numbers indicate averages per configuration. The pure statistical errors
are given in parentheses.
We denote with
$V_{\rm cl}$     - the volume fraction occupied by all topological clusters,
$V_{\rm cl~mon}$ - the volume fraction occupied by clusters containing
                   time-like magnetic monopoles,
$N_{\rm cl}$     - the number of all clusters per configuration,
$N_{\rm cl~mon}$ - the number of clusters containing time-like
                   magnetic monopoles,
$N_{\rm mon}$    - the overall number of dual time-like  links
                   carrying monopole currents,
$N_{\rm mon~cl}$ - the number of dual time-like  links with monopole currents
                   found inside topological clusters,
$N_{\rm loop}$   - the overall number of  thermal monopoles (closed in time direction magnetic current loops) ,
$N_{\rm loop~cl}$ - the number of these thermal monopoles  piercing
                   topological clusters.
}
\end{center}
\end{table*}
\bigskip

\section{Cluster analysis}

An access to the topological charge density at all lattice points at a 
selected time of the gradient flow, allows to perform a cluster analysis of 
the topological density,  with hope to associate the emerging 
clusters with dyons. The method of the analysis was described many times,
for example in~\cite{Bornyakov:2015xao}. As it can be seen from table I, 
clusters occupy several percent of the lattice volume (equal in physical units
to $26~ \rm{fm}^4,~ 5~ \rm{fm}^4,~ 2.5~ \rm{fm}^4$
for ensembles at temperatures $T= 0.79 ~T_c,~ T= 1.27 ~T_c,~ T= 1.5 ~T_c $, respectively) 
and are correlated to magnetic 
monopoles. The correlation is strongest in the confined phase.

We see that at $T=0.79~T_c < T_c$ approximately 100 (from around 300 in total) 
of dual time-like links carrying monopole currents are concentrated in 
only 2.4\% of the lattice occupied by topological clusters. On the other
hand the remaining approximately 200 of dual time-like links carrying 
monopole currents are spreaded over 97.6\% of the rest of the lattice.  
Thus, time-like Abelian magnetic currents are about 20 times more dense 
inside clusters of topological charge than outside. Also, in this phase 
around 10 clusters containing time-like magnetic monopoles at all, are 
pierced totally by around 20=2$\times$10 time-like magnetic current loops 
(of thermal monopoles) as it should be for a KvBLL dyon. This is because 
each dyon is a monopole in one Abelian field and an antimonopole in another 
Abelian field, for the three Abelian fields available in the case of $SU(3)$.

\begin{figure*}[!htb]
\centering
a)\hspace{0.3cm}\includegraphics[width=.45\textwidth]{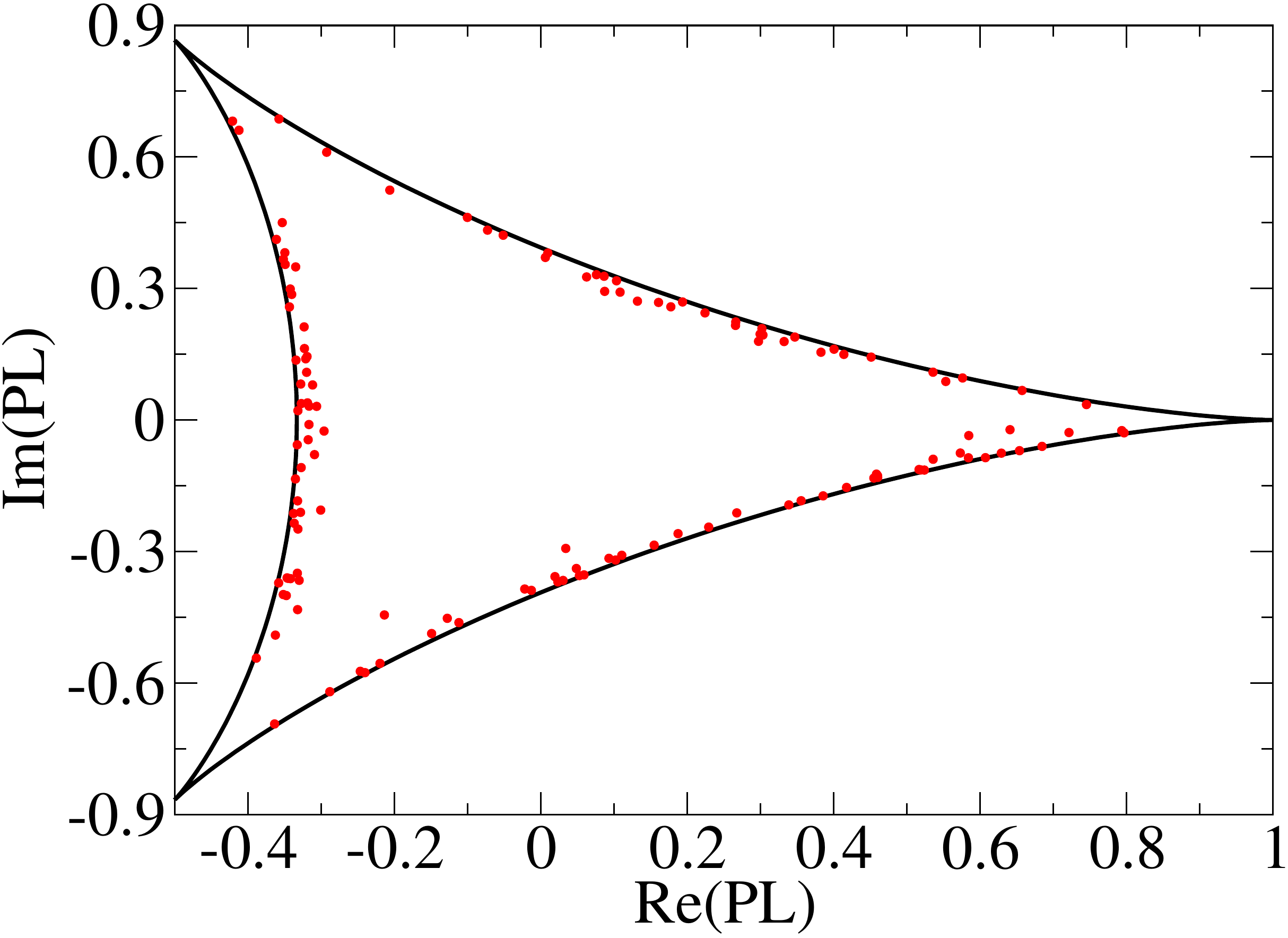}%
\hspace{0.3cm}
b)\hspace{0.3cm}\includegraphics[width=.45\textwidth]{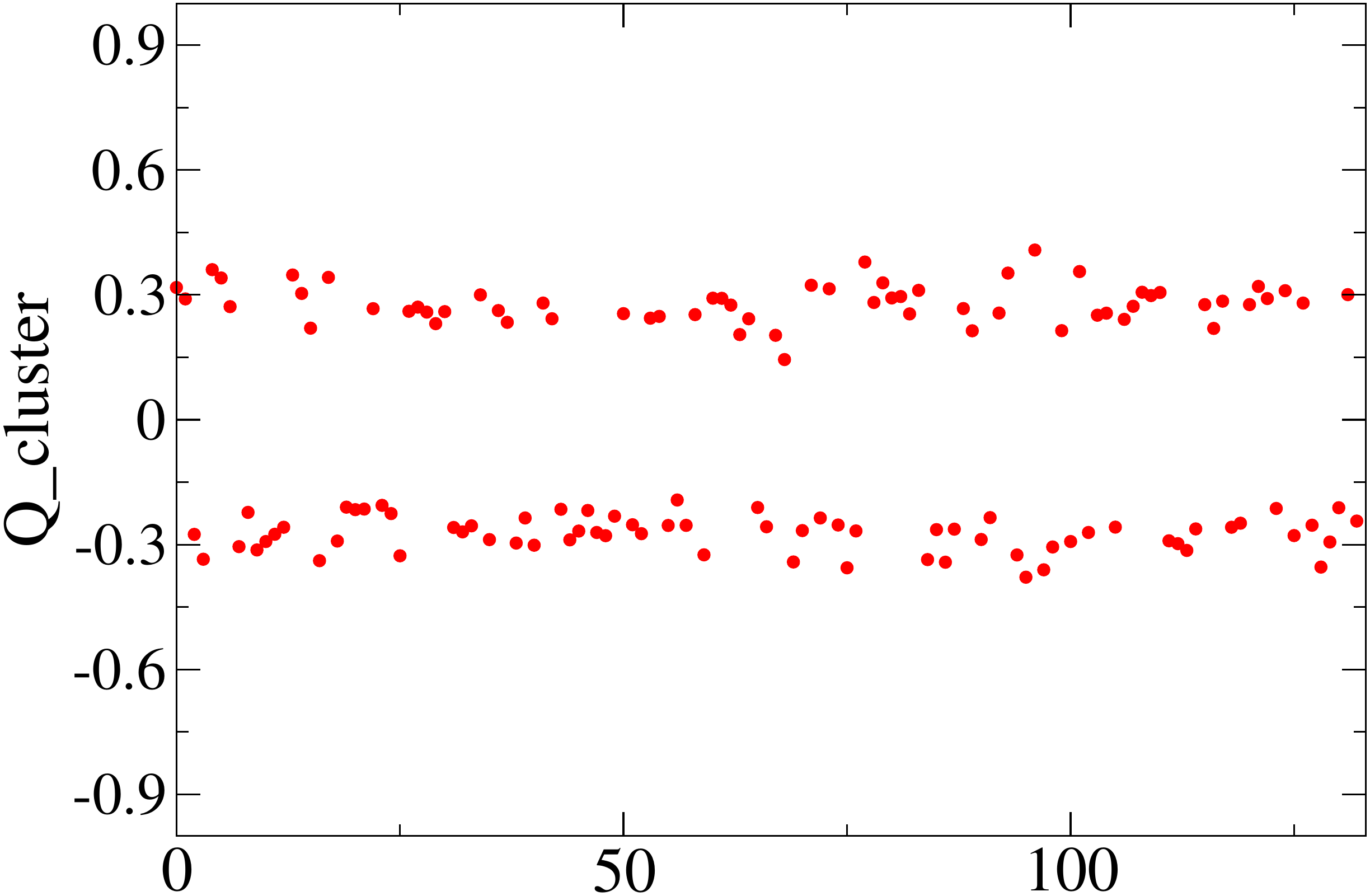}\\
\caption{
a) Scatter plot of the Polyakov loop values at points with 
$\rm{min}(m_1(x), m_2(x), m_3(x))$ for clusters in the confined phase at 
$T= 0.79~T_c$ (shown as circles, only clusters containing  
thermal monopoles are shown),
b) integrated topological charges of these clusters shown also by circles.
Abscissa here shows the number of cluster and for 12 presented configurations there
are $10.4*12\approx 125$ such clusters.}
\label{pl_Q_c}
\end{figure*}

\begin{figure*}[!htb]
\centering
a)\hspace{0.3cm}\includegraphics[width=.45\textwidth]{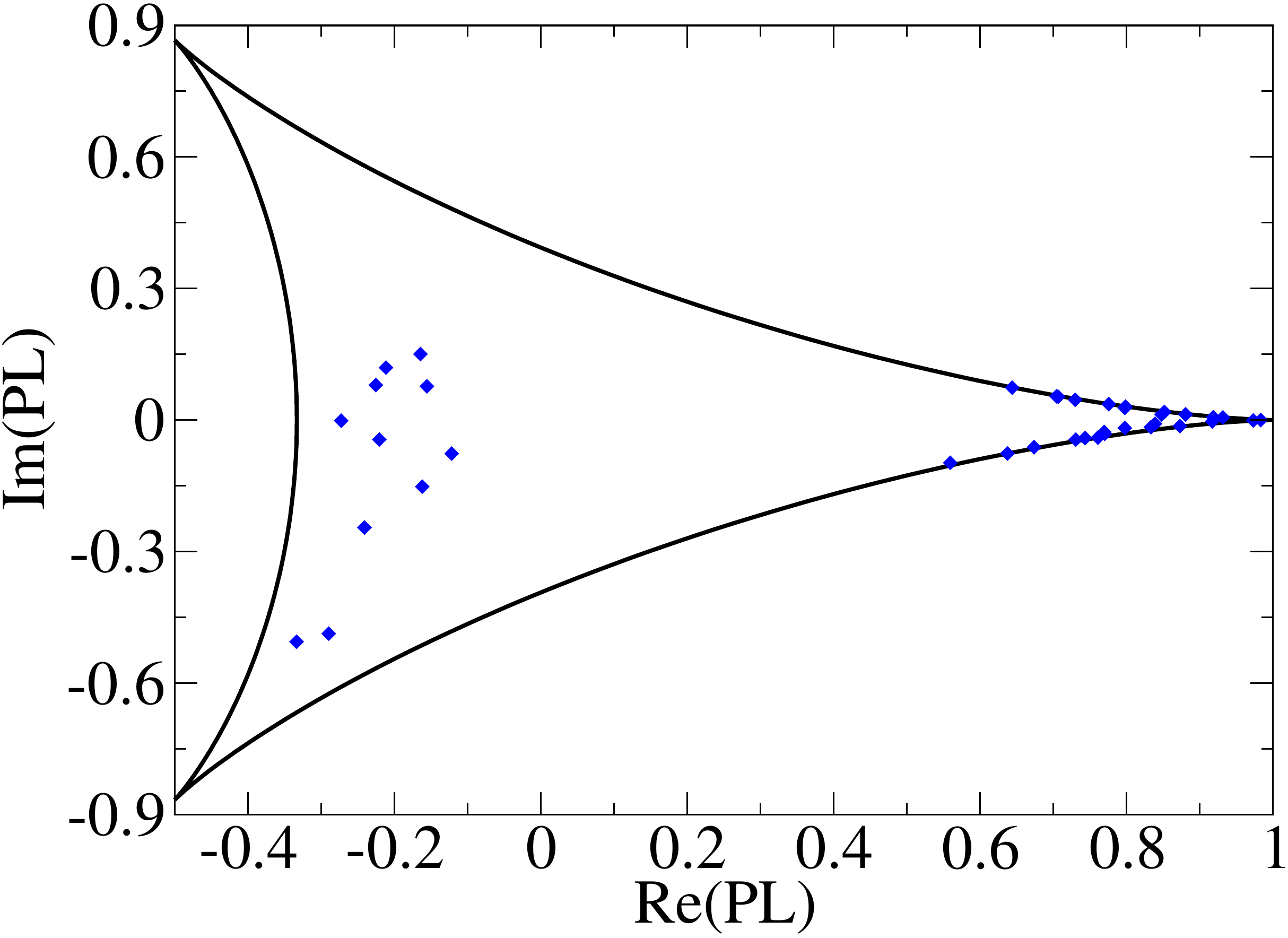}%
\hspace{0.3cm}
b)\hspace{0.3cm}\includegraphics[width=.45\textwidth]{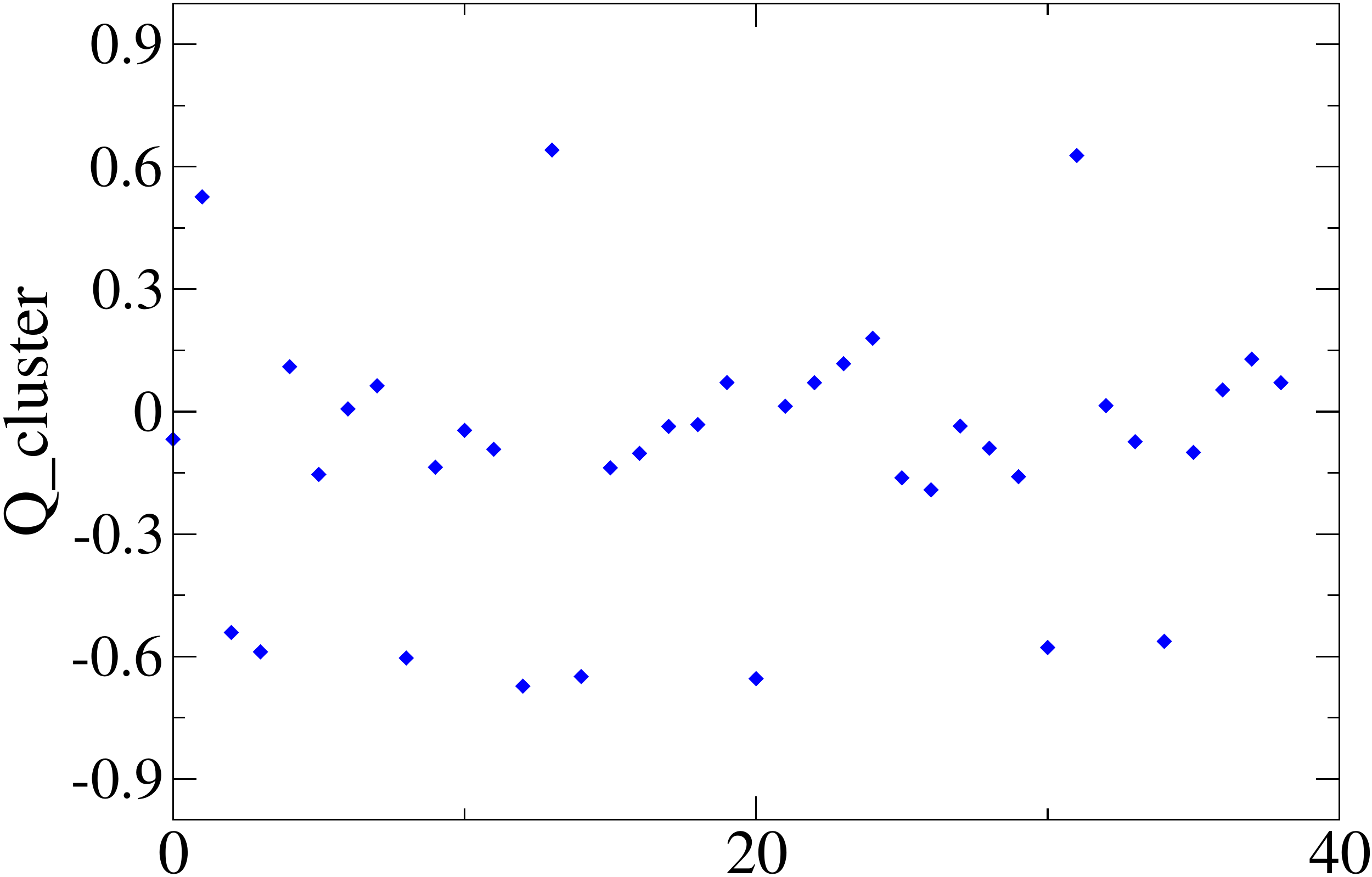}\\
\caption{
a) Scatter plot of the Polyakov loop values at points with 
$\rm{min}(m_1(x), m_2(x), m_3(x))$ for clusters in the deconfined phase 
at $T= 1.27~T_c$ (shown as diamonds, only clusters containing thermal monopoles are shown),
b) integrated topological charges of these clusters shown also by diamonds.
As in Fig.~\ref{pl_Q_c} abscissa shows the number of cluster and for 12 presented configurations there
are $3.2*12 \approx 38$ such clusters.}
\label{pl_Q_d}
\end{figure*}

The (reconstructed) topological charges ($Q_{\rm cl}$) of the clusters containing 
time-like magnetic monopoles are distibuted around $\pm 1/3$ in the confined 
phase (see \Fig{pl_Q_c}b)). In the deconfined phase, these clusters form 
two groups, one of heavy clusters with $0.5 <|Q| < 0.7 $ on one hand 
and another of light clusters with $|Q| < 0.3$ (see \Fig{pl_Q_d}b)) on the 
other hand.

The ``reconstructed'' cluster charge is obtained by summing over the charge 
density according to a procedure described in 
\cite{Ilgenfritz:2004zz,Ilgenfritz:2006ju,Bornyakov:2015xao}, where also the
systematic error is estimated.
 Finally, in the confined phase, the trace of the Polyakov loop measured inside
the clusters in representative points (where the minimum 
$\rm{min}(m_1(x), m_2(x), m_3(x))$ is taken on) is indeed located closely to the 
flanks of the Polyakov triangle. This is an expected feature of KvBLL 
monopole-dyons appearing in three types (of approximately equal charge and
abundance) in the confined phase (see \Fig{pl_Q_c}a)). The three flanks
correspond to the three types of dyon observed.

\begin{table}[ht]
\begin{center}
\vspace*{0.5cm}
\begin{tabular}{|l|l|l|c|c|}
\hline
 phase&$\rho_3(1)$ & $\rho_3(2)$ & $\rho_3(3) $ \\
\hline
$T= 0.79 ~T_c < T_c$&$1.22(2)$ & $1.22(2)$ & $ 1.22(2)$ \\
\hline
$T= 1.27 ~T_c> T_c$&$1.15(7)$ & $1.35(7)$ & $1.35(7) $  \\
\hline
$T= 1.5 ~T_c> T_c$&$0.54(4)$ & $3.1(2)$ & $ 3.1(2) $\\
 \hline
\end{tabular}
 
\label{tabdata1}
\vspace*{0.5cm}
\caption{3d densities $\rho_3(i),~i=1,2,3 $ for heavy, light, light dyons at all three temperatures
in $1/\rm{fm}^3$ units.}
\end{center}
\end{table}

In the deconfined phase, the position of heavy and light clusters 
(represented by the trace of the Polyakov loop) in the Polyakov triangle
is indicated by points with $\rm{min}(m_3(x))$ or $\rm{min}(m_1(x),m_2(x))$ 
for heavy or light clusters (see \Fig{pl_Q_d}a)), respectively. 
We see that 11 out of 38 clusters found in (totally) 12 configurations are heavy
so that heavy dyons are relatively suppressed with respect to the light ones.
This is also an expected feature of KvBLL monopole-dyons.
Finally, we can calculate
3d densities for dyons at all three temperatures (for dyons as static objects 3d densities are
more relevant than 4d densities)
The suppression of heavy dyons
with the increase of temperature is clearly seen (see Table II).


\bigskip

\section{Conclusions}

We have studied the topological structure of $SU(3)$ gluodynamics by 
cluster analysis of the gluonic topological density. 
The gluonic topological charge density was emerging in the process of 
gradient flow with respect to the over-improved action.
Monitoring the IPR of the modulus of the topological density has allowed 
us to stop the gradient flow at the moment when calorons have dissociated 
into dyons due to over-improved character of this process. This has given 
us the possibility to visualize all three dyon constitutents of a 
KvBLL caloron formed in the gluonic field. 
The time-like Abelian monopoles and the specific KvBLL pattern of the 
local holonomy (untraced Polyakov loop) are correlated to topological 
clusters. The reconstructed (summed) values of topological charges for 
each kind of dyons are concentrated near $1/3$ in the confined phase. In the 
deconfined phase, however, the values of the cluster charges 
(characterizing heavy and light dyons) have been found correlated 
with  the local holonomy. The suppression of heavy dyons with the increase of
temperature is clearly seen.

\noi
{\bf Acknowledgments} \\

V.G.B. and B.V.M. are supported by the RFBR grant 16-02-01146a. B.V.M. appreciates the support by the grants RFBR 15-02-07596a.


\end{document}